\documentclass[a4,fleqn,12]{article}

\usepackage{amsmath}             
\usepackage{amssymb}             
\usepackage[english]{babel}    
\usepackage[T1]{fontenc}       
\usepackage[utf8]{inputenc}    
\usepackage{textcomp}          
\usepackage{microtype}           
\usepackage[normalem]{ulem}      
\usepackage{xcolor}              
\usepackage{mathtools}           
\usepackage[symbol]{footmisc}

\usepackage{graphicx}      
\usepackage{multirow}            

\usepackage[margin=1in]{geometry}

\newcommand{\var}{\mbox{var}}

\begin{document}

\begin{center}
{\large The appropriateness of ignorance in the inverse kinetic Ising model }
\end{center}
\begin{flushleft}
\hspace{1in}{\small Benjamin Dunn, Claudia Battistin}
\\
\hspace{1in}{\small Kavli Institute for Systems Neuroscience and Centre for Neural Computation, NTNU}
\\
\hspace{1in}{\small Email: benjamin.dunn@ntnu.no, claudia.battistin@ntnu.no}
\end{flushleft}

  \begin{abstract}
We develop efficient ways to consider and correct 
for the effects of hidden units for the paradigmatic case of the inverse kinetic Ising model with fully asymmetric couplings.
We identify two sources of error in reconstructing the connectivity among the observed units while ignoring part of the network. 
One leads to a systematic bias in the inferred parameters,
whereas the other involves correlations between the visible and hidden populations and has a magnitude that depends on the coupling strength. 
We estimate these two terms using a mean field approach and derive self-consistent equations for the couplings accounting for the systematic bias. 
Through application of these methods on simple networks of varying relative 
population size and connectivity strength, we assess how and under what conditions the hidden portion can 
influence inference and to what degree it can be crudely estimated.
We find that for weak to moderately coupled systems, the effects of the hidden units is a simple rotation that can be easily corrected for. 
For strongly coupled systems, the non-systematic term becomes large and can no longer be safely ignored,
further highlighting the importance of understanding the average strength of couplings for a given system of interest.

  \end{abstract}

\section{Introduction}

Recent technological advances in high-throughput recordings of biological systems 
are enabling the use of statistical tools to ask an entirely new set of questions.
The extent to which a given system can be observed as a whole, however, often limits the applicability of these tools.
For example, currently the most impressive neural recordings are of populations of hundreds or thousands of neurons. 
While a vast improvement from the single recordings from approximately 60 years ago \cite{stevenson2011advances}, 
it is a minuscule fraction of what is likely the relevant population.
Still, with the hope of gaining insight into possible structure of the underlying network, scientists have turned to 
statistical models, such as the generalized linear model (GLM) \cite{nelder1972generalized}, that allow potential \textit{connectivity} (the structure of pairwise interactions and their strength) between
the observed units to be estimated systematically.
Of the many flavors of GLM, the simplest takes the form of a binary spin model or Bernoulli GLM \cite{nelder1972generalized}, 
 also known as the kinetic Ising model \cite{hertz2011ising,roudi2014assembliesXXX} 
in statistical physics. This popular model takes the form,
\begin{equation}\label{eq:transitionIsing}
p(s_i(t+1) | \mathbf{s}(t)) = \frac{\exp(s_i(t+1)H_i(t))}{ 2\cosh(H_i(t))}
\end{equation}
where $s_i(t)=\pm 1$ with $i=1,..,N$ is the state of node $i$ at time $t$. $H_i(t) = \sum_j J_{ij} s_j(t)$ is the input to node $i$, consisting of the field 
generated by the other nodes, with $J_{ij}$ the coupling from node $j$ to $i$. 
It should be noted that at this stage, following (\ref{eq:transitionIsing}), we do not allow for an external field. Even if this model is typically regarded as a model of spiking neurons \cite{coolen2005theory}, since its introduction \cite{little1974existence}, it has been  used for modeling financial markets \cite{zeng2014financial,bouchaud2013crises}. This model is indeed prone to applications in a variety of fields, from biology (e.g. gene interaction networks) to social sciences, where interactions are not symmetric and external inputs are non-stationary. Ultimately the apparent simplicity of the dynamical rule (\ref{eq:transitionIsing}) along with its rich dynamics makes this framework ideally 
suited for developing techniques and gaining insight into the more general class of statistical modeling, and therefore it has recently attracted considerable attention from the statistical physics community \cite{roudi2011dynamical,mezard2011exact,mahmoudi2014generalized,witoelar2011neural,sakellariou2012effect,roudi2011mean,hertz2010inferring}.

When the system is fully observed, the learning of the parameters of this model can be accomplished using techniques that 
take advantage of the convexity of the problem \cite{fahrmeir1985consistency}.
These methods, however, 
are iterative in nature and often require calculating inverses of large matrices with entries consisting of
averages over the data that have to be recomputed at each iteration.
While not a problem for many data sets currently available, the promise of orders of magnitude more data  
would suggest that it will soon be necessary to turn to approximate methods.
\emph{Mean field} approximations \cite{roudi2011mean,roudi2011dynamical,mezard2011exact,mahmoudi2014generalized} that have been developed for this purpose, 
model the fluctuating field $H_i(t)$ with an effective field, reducing a many body problem to many single body problems.
We will focus on fully asymmetric connectivities for which mean field equations, exact in the thermodynamic limit, 
have been first introduced in \cite{mezard2011exact}. 
Here by fully asymmetric couplings we mean fully asymmetric Sherrington-Kirkpatrick  couplings \cite{sherrington1975solvable}: the $J_{ij}$s are assumed to be independently drawn ($J_{ij}$ statistically independent from $J_{ji}$) from a normal distribution with mean $J_1/N$ and
standard deviation $J_0/\sqrt{N}$. 
For large asymmetric networks the authors in \cite{mezard2011exact} approach the inverse problem deriving a relation between correlations and couplings:
 \begin{equation} \label{eq:meanfieldequations}
   \mathbf{D}(t) = \mathbf{A}(t)\mathbf{J}\mathbf{C}(t)
 \end{equation}
where $D(t)$/$C(t)$ are the time-delayed and equal-time correlation matrices
 \begin{eqnarray}
   &&   D_{ij}(t) = \langle ds_i(t) ds_j(t-1) \rangle, \label{eq:DefeqCorr}\\
   &&   C_{ij}(t) = \langle ds_i(t)ds_j(t) \rangle.
 \end{eqnarray}
Here $\langle \cdot\rangle$ indicates averaging over the distribution ${\rm P}(\mathbf{s})=\prod_{i,t}p(s_i(t+1) | \mathbf{s}(t))$, where $p(s_i(t+1) | \mathbf{s}(t))$ is defined in (\ref{eq:transitionIsing}), while $ds_i(t)\equiv s_i(t)-m_i(t)$, where $m_i(t)=\langle s_i(t)\rangle $.
In  (\ref{eq:meanfieldequations}) the diagonal matrix $\mathbf{A}$ is: 
\begin{equation} \label{eq:AF} 
  A_{ij}(t) = \delta_{ij} \int Dx \big[ 1 - \tanh^2 \big(g_i(t) + x \sqrt{\Delta_i(t)} \big) \big]
\end{equation}
where $\Delta_i(t) = \sum\nolimits_{k}J_{ik}^2(1-m_k(t)^2)$, $Dx = \frac{dx}{\sqrt{2 \pi}} e^{-x^2/2}$ and $g_i(t)=\langle H_i(t)\rangle$. 

Interestingly, this relationship has been shown to be exact for fully observed systems with asymmetric couplings, even in the strongly coupled regime \cite{mezard2011exact}.
It is, however, very uncommon that a system is able to be fully observed, necessitating further understanding of
what kind of errors we incur by ignoring the hidden population and perhaps develop approximations to account for them more explicitly.
By explaining away correlations due to these unobserved but potentially relevant features, we begin to address
the concerns we share with many about how informative are observed data and perhaps, some day, make stronger claims about our abilities to uncover actual network features.

Besides its potential applications, the hidden population problem in the KI model is intriguing both from computational and ``physical'' points of view. Computationally, since exact and Monte Carlo sampling of the states space are expensive if not unfeasible; physically, since the likelihood of the data  in partially observed systems, as opposed to the fully observed ones, is not convex. For GLMs this problem has been approached previously by, for example, assuming the hidden units 
provide a source of Gaussian random noise, as in \cite{kulkarni2007common}, and
by approximate \textit{Expectation Maximization} (EM) algorithms as in \cite{pillow2007neural,dunn2013learning,tyrcha2014network,battistin2015belief}; see also \cite{bachschmid2014inferring}. 
In this framework \cite{sundberg1974maximum,dempster1977maximum}, the likelihood of the observed time series is maximized by alternatively estimating 
expected values of the joint (observed + hidden) distribution at fixed connectivity and then updating the couplings accordingly. 
When the hidden population constitutes a minor component of the entire network the EM algorithm has been proven accurate \cite{wu1983convergence,mclachlan2007algorithm}, while in the highly subsampled regime the problem becomes burdened with local optima. 

\begin{figure} 
\centering
\resizebox{170mm}{!}{
\includegraphics{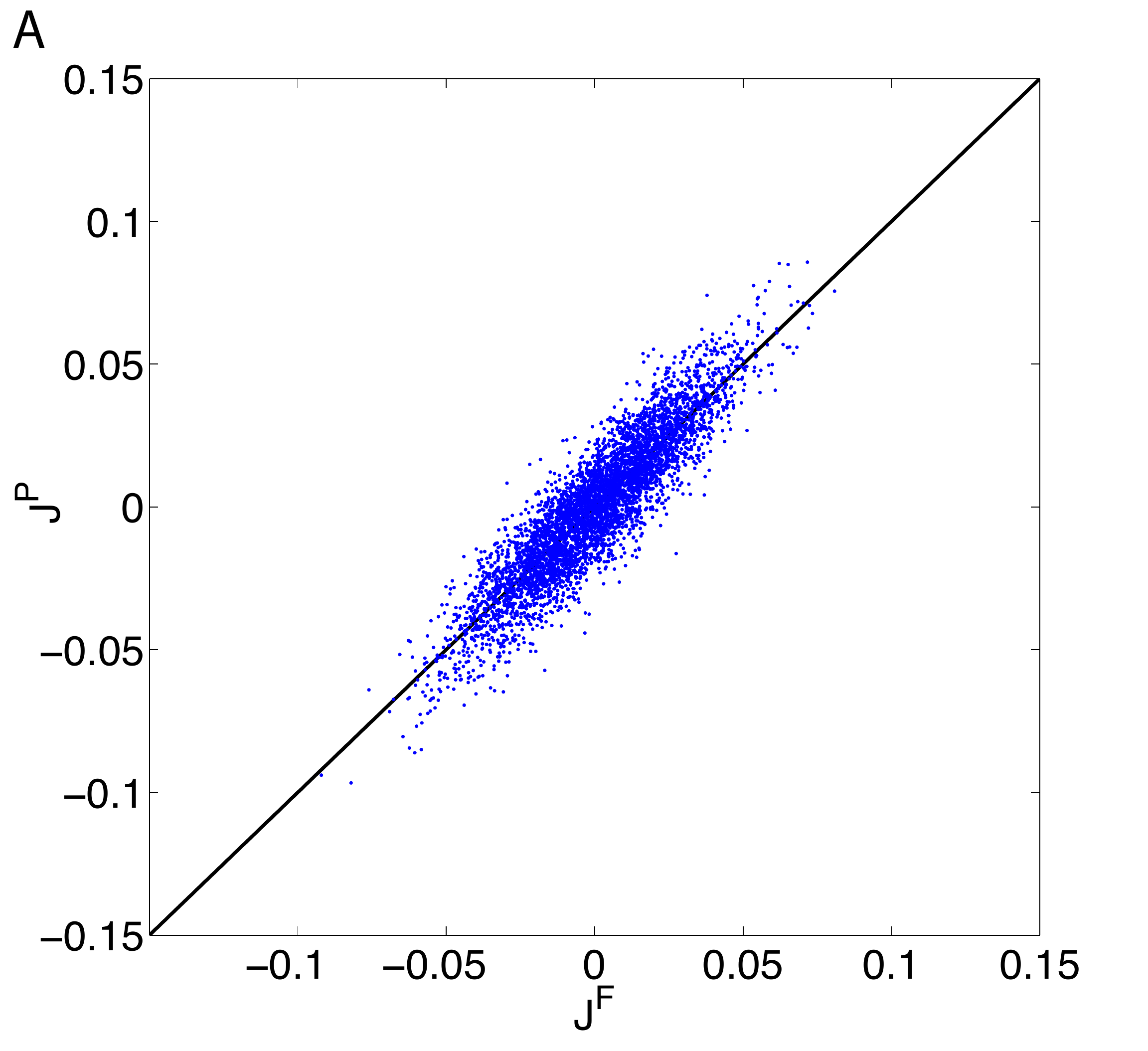}
\includegraphics{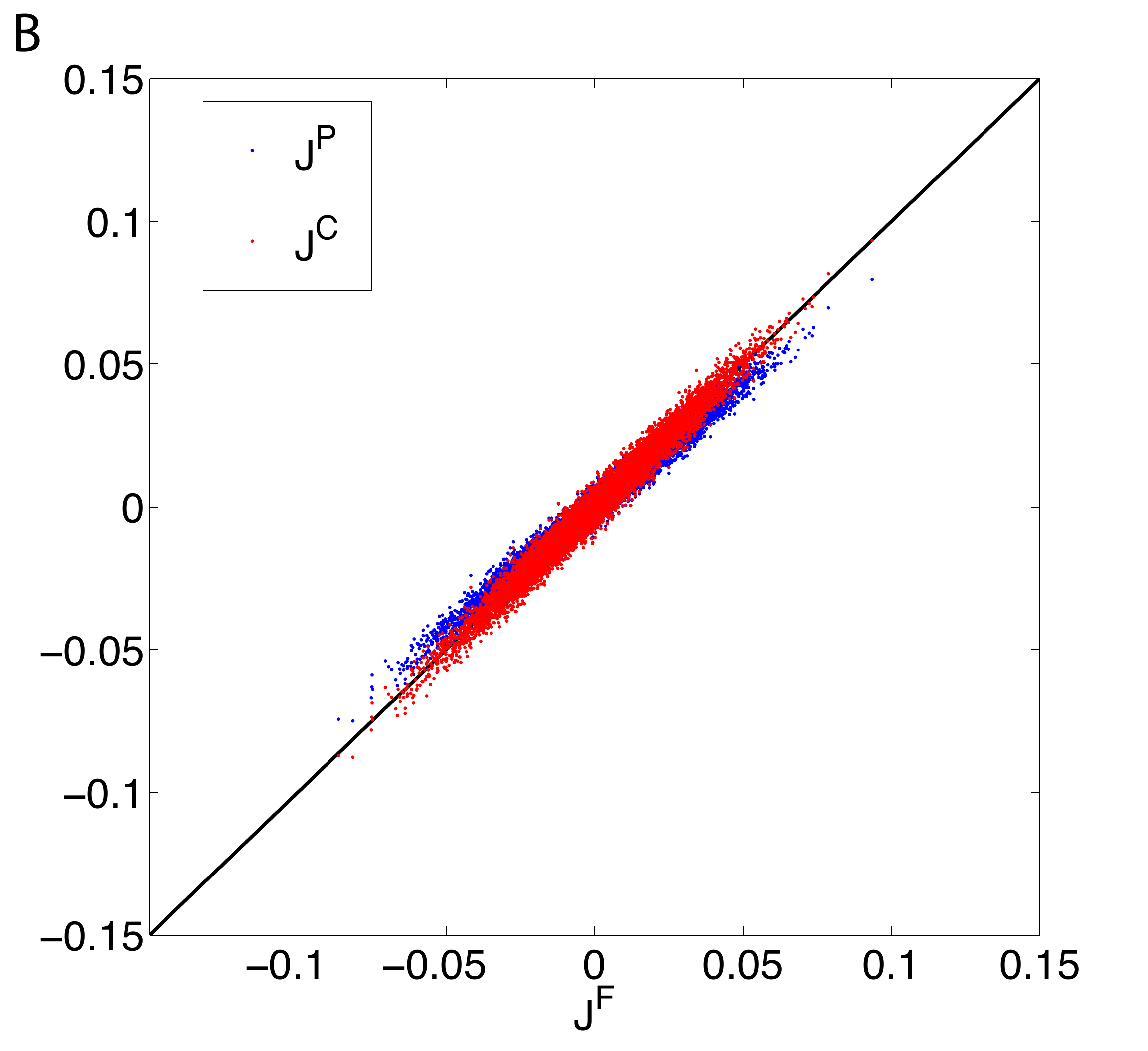}  }
\caption{ \label{fig:Scatter}Network reconstruction in presence of hidden nodes. Scatter plot of the inferred couplings among $P=100$ nodes out of $N=500$: observing only the $P$ nodes $\mathbf{J^P}$ versus using data from the whole network $\mathbf{J^F}$. SK couplings $\mathbf{J}$ independently drawn from a normal distribution with zero mean and standard deviation $J_0/\sqrt{N}$ with $J_0=0.5$; data length $L=10^7$ $^\dag$.  \textbf{A}: equilibrium Ising model, symmetrized version of the $\mathbf{J}$, inference performed using the Na\"{i}ve Mean Field algorithm \cite{tanaka1998mean}\cite{kappen1998efficient}. \textbf{B}: kinetic Ising model, fully asymmetric couplings $\mathbf{J}$; inference performed using the mean field learning from \cite{mezard2011exact} (blue)---not visually distinguishable from those learned via exact learning \cite{hertz2011ising}---, corrected using (\ref{eq:rotationequation}) (red). 
}
\end{figure}

There are few cases where even in the presence of many hidden units, the connected and
disconnected pairs, can be well identified from the inferred connections. For instance, 
\cite{lezon2006using} reported that increasing the number of genes used in their inference did not significantly
change the important couplings between the observed nodes inferred using a maximum
entropy model. In this specific case the insensitivity of the reconstruction to the inclusion of further nodes seems to follow from the 
specific choice of the inclusion criterion, such that the presence of the additional nodes does not affect significantly the correlations between the original ones. 
A second example regards a subpopulation of neurons: in \cite{roudi2009ising} the connectivity exhibits a rescaling but no remarkable restructuring by 
observing larger and larger populations. This effect has been observed even when the generative model is an equilibrium model, 
as in \cite{haiping2015effects} and can be attributed to the mean field properties of the underlying network, 
in a way that the unobserved neurons affect the correlations among the observed neurons as a whole. 
These results suggest that, when mean field arguments apply, the inference of the couplings between observed nodes can be corrected 
for the unobserved through a global transformation. In figure \ref{fig:Scatter} A-B we show that the scaling of the couplings is not an effect limited to 
equilibrium models but it extends to the kinetic model with asymmetric couplings and does not rely on approximate learning. 

In this paper we follow up on these ideas analyzing the effects of the hidden variables on the reconstruction of the couplings when you only see $P$ neurons ($\mathbf{J}^P$), comparing them with those inferred from the full network ($\mathbf{J}^F$), that we will partition as follows:  
\begin{equation}
\mathbf{J^F }= \left[  
\begin{smallmatrix} 
\mathbf{J_{ss}^F} & \mathbf{J_{s \sigma}^F} \\ 
\mathbf{J_{\sigma s}^F} & \mathbf{J_{\sigma \sigma}^F}  
\end{smallmatrix} \right] 
\label{eq:Jblocks}
\end{equation}
where $\mathbf{s}$ indicates the observed nodes, while $\boldsymbol{\sigma}$ the hidden ones. Exploiting (\ref{eq:meanfieldequations}) the difference between the inferred couplings can be written as: 
\begin{equation} \label{eq:approximationerror}
  \mathbf{J^P} - \mathbf{J_{ss}^F} =\boldsymbol{\Xi}+\boldsymbol{\Gamma}
\end{equation}
with  
\begin{eqnarray}
\boldsymbol{\Xi}&\equiv & \left( \mathbf{I} - \left[\mathbf{A_{ss}^F}\right]^{-1}\mathbf{A^P} \right) \mathbf{J^P} \label{eq:DefXi}\\
 \boldsymbol{\Gamma}&\equiv &\mathbf{J^F_{s \sigma}}\mathbf{C_{\sigma s}} \left[\mathbf{C_{ss}}\right]^{-1}\label{eq:DefGamma} 
\end{eqnarray}
where the matrix $\mathbf{A}$  inherits the label, $\mathbf{F}$ or $\mathbf{P}$, from the couplings that enter its expression and $\mathbf{A^F}$ is partitioned, just like $\mathbf{C}$, accordingly to (\ref{eq:Jblocks}). For sake of clarity from  (\ref{eq:approximationerror}) on we drop the time indices, bearing in mind that the inference problem can be solved in parallel at each time step. The following derivation will involve dynamical quantities  at fixed  time $t$, such that, for example, $m_i$ will now indicate $m_i(t)$ instead.
  
Thus the error can be seen as having two components, a systematic term $\boldsymbol{\Xi}$ and a non-systematic one $\boldsymbol{\Gamma}$ that is shaped by the observed to observed and hidden to observed correlations. 

In the next sections we seek to find a simple way to deal with the hidden population by estimating these two contributions in a mean-field type approach.

\renewcommand{\thefootnote}{\fnsymbol{footnote}}

\footnotetext[2]{In the regime of large $L$ and in absence of an external field, we consider the process as being stationary, such that, when solving (\ref{eq:meanfieldequations}) we replace population averages $<...>$ with time averages. } 

\section{The systematic error term $\boldsymbol{\Xi}$}

In this section we are going to study the error term $\boldsymbol{\Xi}$ in (\ref{eq:approximationerror}). It generates a systematic bias in the learned couplings, that affects concertedly their magnitude, as shown in Figure \ref{fig:Scatter}.

We will now evaluate $\boldsymbol{\Xi}$ using a mean field approximation, building up on the assumption that statistically the hidden part of the network 
differs from the observed one neither architecturally nor dynamically. 
The local field on the observed node $i$ can then be expressed as the mean field generated by the observed nodes plus a random variable $\rho$:
\begin{equation}
H_i=\sum\nolimits_j^P J_{ij}m_j +\rho_i
\end{equation}
For zero mean couplings ($J_1=0$) and fields, we have $\langle {\rho_i} \rangle = 0$ and we approximate its variance with
\begin{equation} 
\var(\rho_i) \approx \sum\nolimits_{k}^P(J_{ik})^2(1-m_k^2)+(N-P)\var\left( \mathbf{J}\right) \left( 1-\overline{ \mathbf{m^2}} \right)
\label{eq:varRho}.
\end{equation}
where $\overline{\cdot}$ and $\var\left(\cdot\right)$ indicate respectively mean and variance over the space of the observed variables, namely $\overline{\mathbf{m^2}}\equiv\frac{1}{P}\sum_j m_j^2$ and $\var\left( \mathbf{J}\right)=\frac{1}{P^2}\sum_{ij}\left(J_{ij}-\overline{\mathbf{J}}\right)^2$. Notice that a non-zero mean of the couplings (e.g. $J_1/N$) would only affect the mean of the variable $\rho$, resulting into an additional term that, in the mean field framework can be approximated as $(N-P)\overline{\mathbf{J}}\overline{\mathbf{m}}$. 

In the large systems limit, being the couplings $J_{ij}$ iid random variables of order $1/\sqrt{N}$, the Central Limit theorem applies to the fields $H_i(t)$. Thus $\rho$ is a Gaussian distributed random variable, such that $A^F_{ij}$ in (\ref{eq:approximationerror}) becomes
\begin{equation} \label{eq:AFapprox}
  \hat{A}^F_{ij} = \delta_{ij} \int Dx \Big[ 1 - \tanh^2 \Big(\sum\nolimits_{k=1}^P J_{ik}^F m_k + x (\var(\rho_i^F) )^{\frac{1}{2}} \Big) \Big] 
\end{equation}
with $\var(\rho_i^F)$ being the variance in (\ref{eq:varRho}) with couplings $\mathbf{J^F_{ss}}$ inferred using the full data set.

Finally we can substitute (\ref{eq:AFapprox}) in the expression for the systematic component of the error:
\begin{equation} 
\boldsymbol{\Xi} \approx \big( \mathbf{I} - \big[\mathbf{\hat{A}_{ss}^F}\big]^{-1}\mathbf{A^P} \big) \mathbf{J^P}.
\label{eq:XiApprox}
\end{equation}

From (\ref{eq:approximationerror}) we can construct
a set of self-consistent equation for the couplings that would be inferred using the full data set, $\mathbf{\hat{J}^F_{ss}}$,
accounting for this systematic error on the couplings inferred from the reduced population, thus
\begin{equation} 
  \mathbf{\hat{J}^F_{ss}} = \big[\mathbf{\hat{A}_{ss}^F}\big]^{-1}\mathbf{A^P} \mathbf{J^P} - \boldsymbol{\Gamma} 
  \label{eq:SelfConsistentJXi}
\end{equation}
where in $\mathbf{\hat{A}_{ss}^F}$ we have replaced $\mathbf{{J}_{ss}^F}$ with $\mathbf{\hat{J}_{ss}^F}$.

Within the limits of applicability of the mean field approximation developed in this section, we can now explain why the global structure of the weights is retained when ignoring the systematic error term in the reconstruction. Despite (\ref{eq:SelfConsistentJXi}) is not an explicit relation between $\mathbf{J^F_{ss}}$ and $\mathbf{J^P}$, we expect $\mathbf{\hat{A}_{ss}^F}$ to be  weakly sensitive to the identity of the couplings that enter (\ref{eq:AFapprox}), when both couplings and states have zero mean. If this holds then the main difference between $\mathbf{\hat{A}_{ss}^F}$ and $\mathbf{A^P}$  in (\ref{eq:SelfConsistentJXi}), consist of a uniform shift in the variance of the gaussian integral associated with (\ref{eq:AFapprox}) given by the second term in (\ref{eq:varRho}). This produces a global rescaling of the couplings reflected in the rotation exemplified in Figure \ref{fig:Scatter}C.

\vspace{\baselineskip}

We can now turn our attention to the second error term.

\section{The non-systematic error term $\boldsymbol{\Gamma}$}

The component $\boldsymbol{\Gamma}$ of the error in (\ref{eq:approximationerror}) can be interpreted as a propagation of the correlations between observed and hidden variables on the correlations between observed units via the hidden to observed interactions. Notice that this term indeed vanishes when there is no hidden to observed coupling, even though the two populations might covary.

For couplings with a mean and standard deviation of zero ($J_1=0$) and $J_0/\sqrt{N}$, respectively, we have average $\bar{\boldsymbol{\Gamma}} = 0$. Thus in order to quantify the extent to which this error term affects the learned couplings one has to estimate its variance  $\var\left(\boldsymbol{\Gamma}\right)$. Since we note that for
small $J_0$, the terms on the diagonal of $\left[\mathbf{C_{ss}}\right]^{-1}$ dominate 
\begin{equation} 
  \var\left(\boldsymbol{\Gamma}\right) \approx \frac{N-P}{N}J_0^2\var\left(\mathbf{C^{\mbox{off}}}\right)\\ \label{eq:gammaApprox}
\end{equation}

where $\mathbf{C^{\mbox{off}}}$ are the off-diagonal elements of $\mathbf{C}$. We can approximate $\mathbf{C^{\mbox{off}}}$ by observing, as in \cite{mezard2011exact}, 
that the joint distribution of the fluctuations around the mean local field to each spin $\delta g_i=\sum_l J_{il}\delta s_l$ is $\epsilon_{ij}=\langle \delta g_i \delta g_j \rangle = \left[ \mathbf{JCJ}^T \right]_{ij}$.
This leads to a joint distribution of $x=\delta g_i$ and $y =\delta g_j$ of the form \cite{mezard2011exact}
\begin{equation}
  P(x,y) = \frac{1}{2 \pi \sqrt{\Delta_i \Delta_j}} \exp \left( - \frac{x^2}{2 \Delta_i} - \frac{y^2}{2 \Delta_j} + \epsilon_{ij} \frac{xy}{\Delta_i \Delta_j} \right)
  \label{eq:jointDeltag}
\end{equation}
Then equal-time correlations (\ref{eq:DefeqCorr}) can be written as:
\begin{align} \nonumber
  C_{ij} &= \int dx dy P(x,y) \delta \tanh(h_i + g_i + x) \delta \tanh(h_j + g_j + y) \\ \nonumber
         &\approx \frac{\epsilon_{ij}}{\Delta_i \Delta_j} \int \frac{dx}{\sqrt{2 \pi \Delta_i}} 
                   \frac{dy}{\sqrt{2 \pi \Delta_j}} \exp{\left[-\frac{x^2}{2 \Delta_i}-
                   \frac{y^2}{2 \Delta_j}\right]} xy \delta \tanh(h_i + g_i + x) \delta \tanh(h_j + g_j + y) \\ \label{eq:eqCorr}
         &= \epsilon_{ij} a_{i} a_{j} 
\end{align}
where $A_{ij}=\delta_{ij}a_i$ is defined in (\ref{eq:AF}).
In (\ref{eq:eqCorr}) we have first expanded the exponential in the first line to the linear order in 
$\epsilon$ (recall that $\epsilon \propto N^{-1/2}$), then we performed the integration by parts to get to the third line.
Using (\ref{eq:eqCorr}) one can show that the variance of the off-diagonal elements of $\mathbf{C}$ is modulated by the couplings strength $J_0$
\begin{equation} 
  \var\left(\mathbf{C^{\mbox{off}}}\right)\approx \frac{J_0^4 \var\left(\mathbf{a}\right
  )^2 \sum_j(1-m_j^2)^2}{N^2[1-J_0^4 \var\left(\mathbf{a}\right)^2]} \label{eq:varCoff}.
\end{equation}
where $\mathbf{a}$ is the vector of the diagonal entries of $\mathbf{A}$ defined in (\ref{eq:AF}).

Replacing (\ref{eq:gammaApprox}) in (\ref{eq:varCoff}) we obtain an estimate for the variance of the non-systematic component of the error in (\ref{eq:approximationerror}):
\begin{eqnarray} 
&& \var (\boldsymbol{\Gamma}) \approx \frac{J_0^6(N-P) \var(\mathbf{a})^2 \sum_j(1-m_j^2)^2}{N^3[1-J_0^4 \var(\mathbf{a})^2]} \\ \label{eq:simpleapproxfornonsys}
&& \hspace{5mm} \approx J_0^6(N-P)/N^2 
\end{eqnarray}
where in the second line we note that, for small values of $J_0$, one has $A_{ij} \approx \delta_{ij} (1-m_i^2)$.
From (\ref{eq:simpleapproxfornonsys}) one can identify two limiting behaviors of the non-systematic error term when compared with values of the inferred couplings, for large systems with weak interactions.
If the fraction of hidden units is large ($\frac{N-P}{N}\sim O(1)$), the relative error $\var(\mathbf{\Gamma})/\var(\mathbf{J})\sim J_0^4$  stays finite even in the regime of perfect sampling; while it vanishes if the fraction of hidden units is small. Notice that a non-zero mean of the couplings (e.g. $J_1/N$) won't change our estimate of $\var(\boldsymbol{\Gamma})$ to the leading order.

\begin{figure} 
\centering
\resizebox{110mm}{!}{\includegraphics{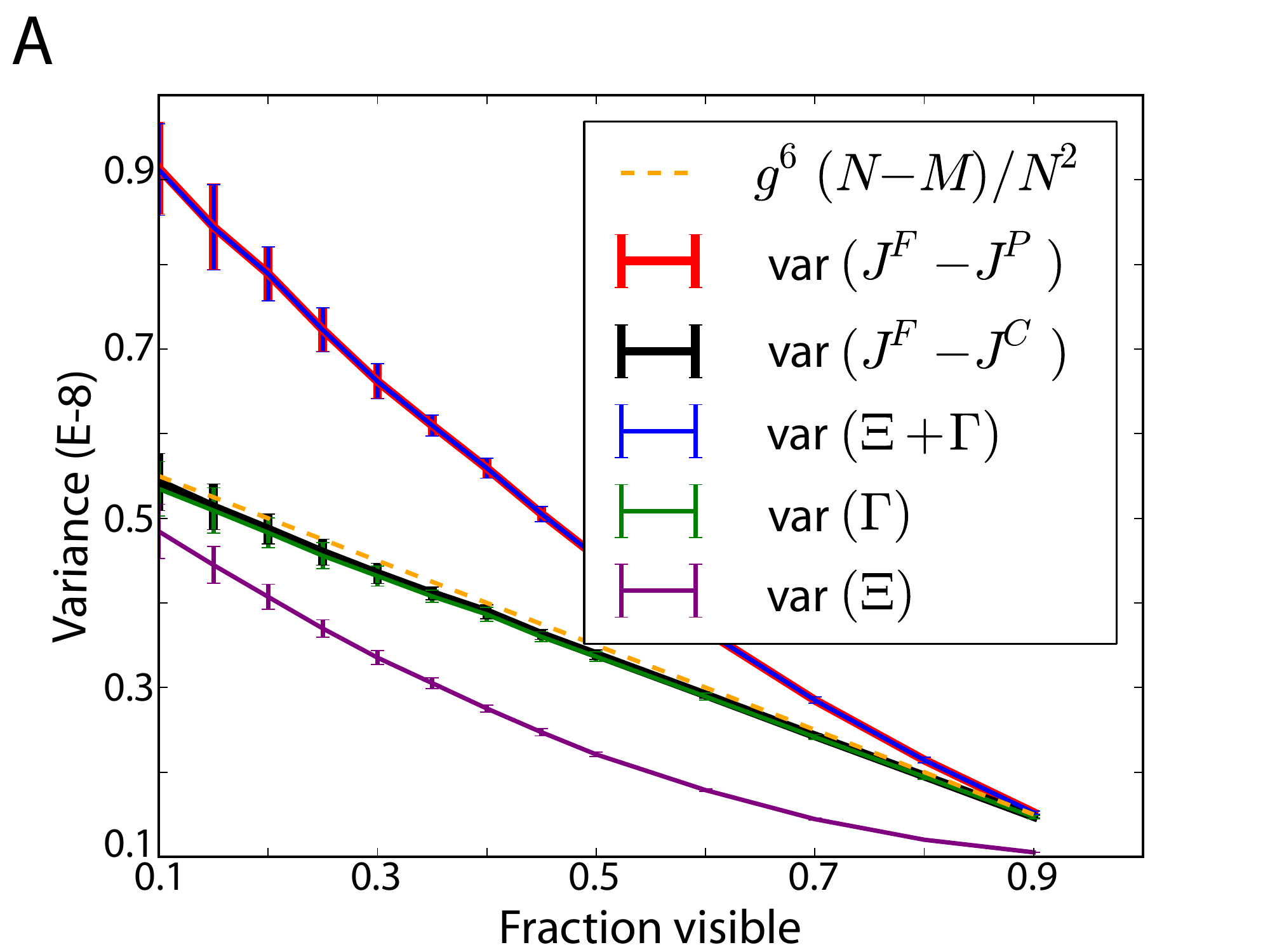}}
\resizebox{170mm}{!}{\includegraphics{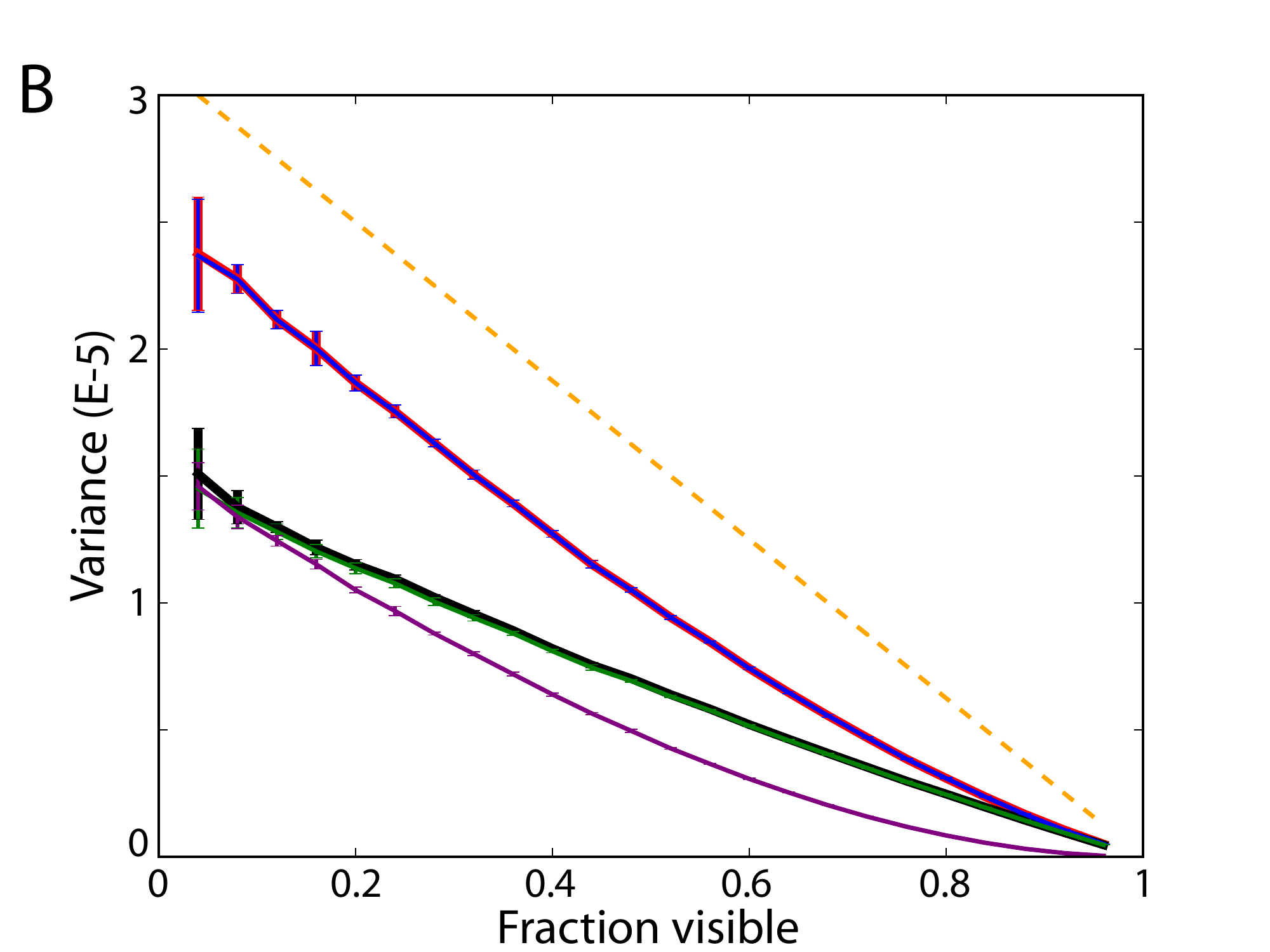}\includegraphics{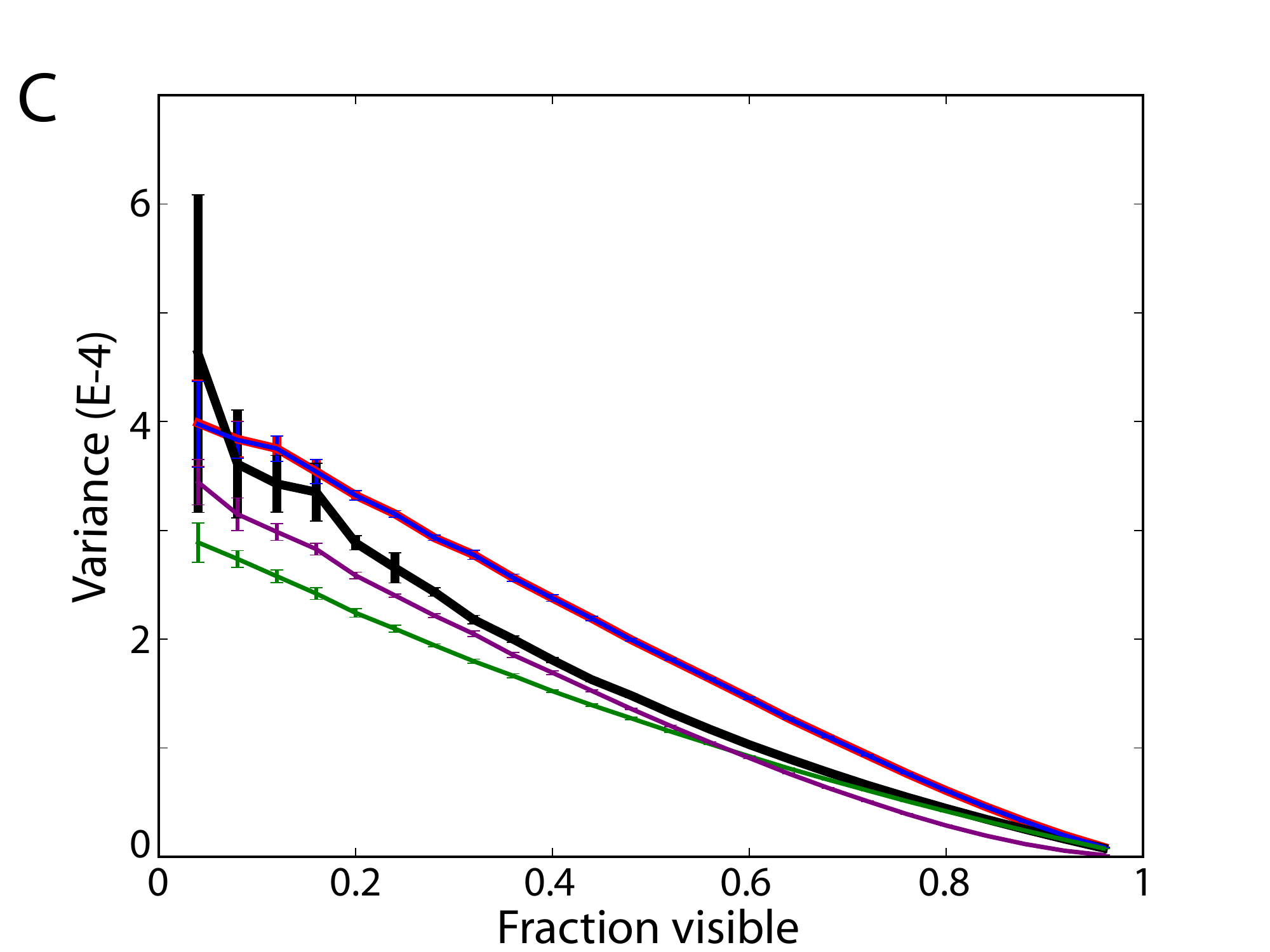}}
\caption{\label{fig:rotationresults} Correcting for the unobserved. 
The systematic bias, $\boldsymbol{\Xi}$, is partially removed by solving the self-consistent equation \ref{eq:rotationequation}.
The resulting corrected couplings, $\mathbf{J^C}$, provide a more accurate estimation of the ground truth with the remaining
error estimated by $\var(\boldsymbol{\Gamma})$, i.e. 
$\var({\mathbf{J^F}-\mathbf{J^P}}) \approx \var(\boldsymbol{\Xi}+\boldsymbol{\Gamma})$ 
and $\var(\mathbf{J^F} - \mathbf{J^C}) \approx \var(\boldsymbol{\Gamma})$.
(\textbf{A}) For weak couplings, $J_0=0.1$, the approximations developed here perform very well, as expected.
For stronger couplings $J_0=0.5$ (\textbf{B}) and $J_0=1$ (\textbf{C}) the corrected couplings 
do result in an improved reconstruction in every case except for the strongly coupled and heavily subsampled regime.
Each inference was done 50 times, randomly selecting hidden nodes each time, and the resulting variances averaged. 
Error bars represent one standard deviation. 
The data was generated from a network with 200 nodes and a data length $L = 1.5\times10^{10}$ $^\dag$ for $J_0=0.1$ (\textbf{A})
and 500 neurons with $L = 10^7$ $^\dag$, for $J_0= 0.5, 1$ (\textbf{B}-\textbf{C}).}
\end{figure}

Numerical simulations (Figure \ref{fig:rotationresults}) confirm that for weak couplings, $J_0=0.1$ (Figure \ref{fig:rotationresults}A)
equation (\ref{eq:simpleapproxfornonsys}) provides a very accurate estimate of this error term while for stronger couplings $J_0=0.5$ (Figure \ref{fig:rotationresults}B)
this has already diverged. The approximation was not included for $J_0=1$ given that it lies outside the bounds of the figure.

\begin{figure} 
\centering
\resizebox{170mm}{!}{\includegraphics{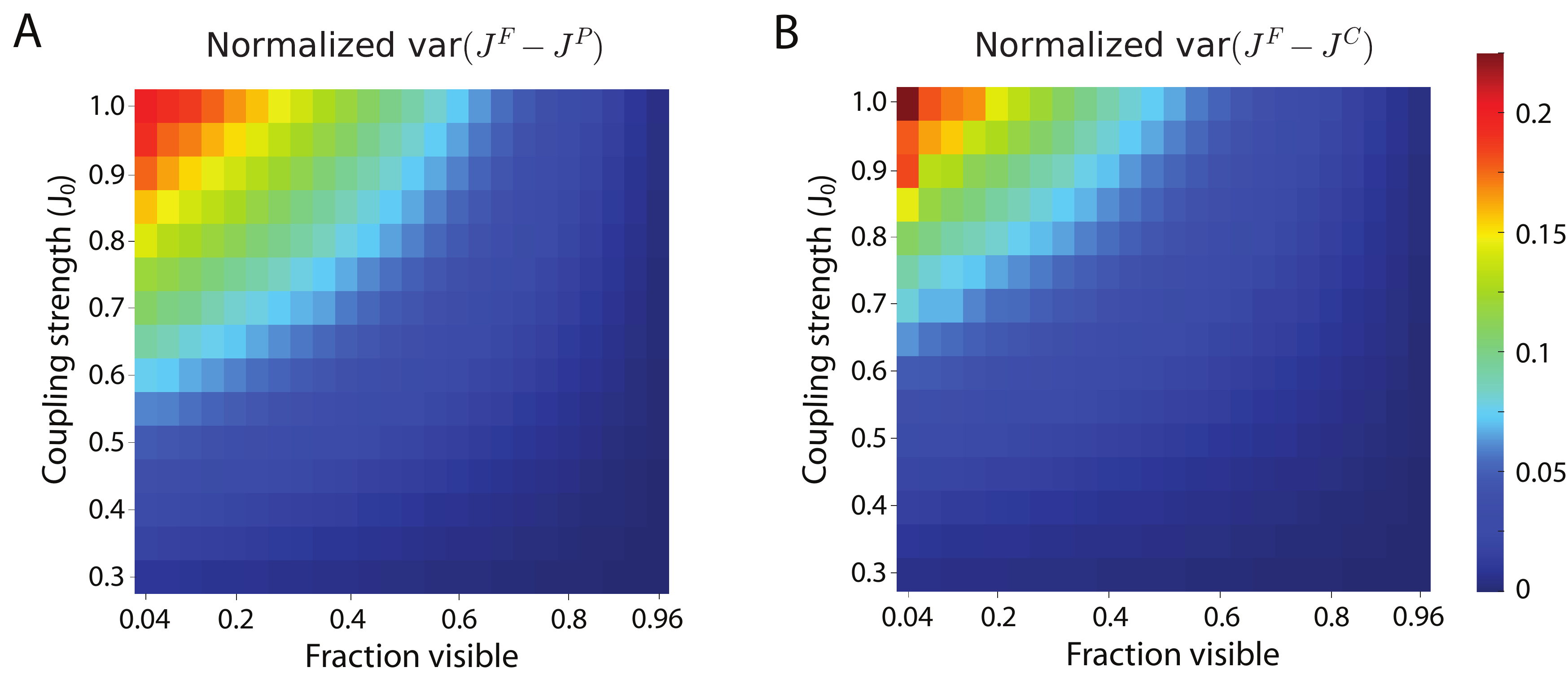}}
\caption{\label{fig:phasedaigrams} Normalized relative errors for the couplings inferred from the partially 
  observed system (\textbf{A}) and using the approach described in equation (\ref{eq:rotationequation}) (\textbf{B}) as a function of subsampling (x-axis)
  and couplings strength $J_0$ (y-axis). Here the error terms are averages over 50 random selections of the hidden nodes as in Figure \ref{fig:rotationresults}, normalized
  by the variance of the couplings ($J_0^2/N$). It is interesting to note that the simple correction is able to significantly
  improve the inference for all but the strongly-coupled, highly-subsampled regime.
}
\end{figure}

\begin{figure} 
\centering
\resizebox{110mm}{!}{\includegraphics{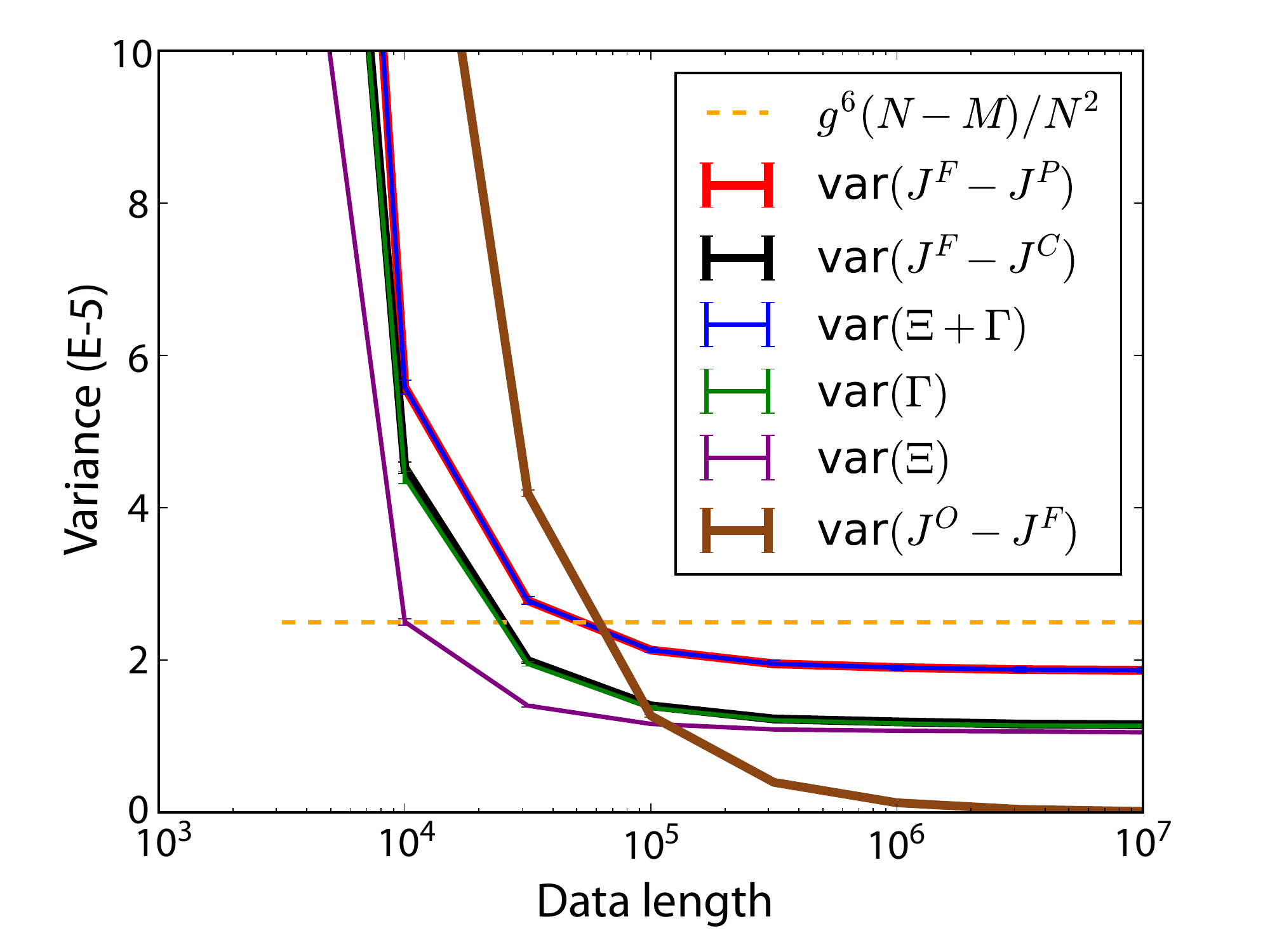}}
\caption{\label{fig:undersampling} Hidden nodes problem as a function of data length. 
  With increasing data length, the estimates and error terms from Figure \ref{fig:rotationresults}
  continue to decrease until saturation.
  It is clear that even with infinite sampling the resulting errors due to subsampling would persist while 
  for the fully observed system (brown) the overall error tends to zero, as one would expect.
  It is important to note that the problem of subsampling quickly becomes relevant, with, 
  in this example,  the error due to subsampling nodes surpassing the error due to data length at below $10^5$.
  In this example, $J_0=0.5$ with a visible fraction of $0.2$.}
\end{figure}

Finally, for small values of $J_0$, we can assume the non-systematic term $\pmb{\Gamma}$ term is negligible
and approximate the couplings using the self-consistent equation
\begin{equation} \label{eq:rotationequation}
  \mathbf{J^C} = \big[\mathbf{A^C}\big]^{-1}\mathbf{A^P} \mathbf{J^P} 
\end{equation}
where $\mathbf{A^C}$ is defined as $\hat{\mathbf{A}}^F$ in (\ref{eq:AFapprox}) but replacing $\mathbf{J^F}$ with $\mathbf{J^C}$. Figure \ref{fig:Scatter}C shows an example of the correction to the inferred couplings from the full data set provided by (\ref{eq:rotationequation}), while Figure \ref{fig:rotationresults} demonstrates that using (\ref{eq:rotationequation})
improves the learning systematically reducing the error in 
some cases to nearly half.
Interestingly, the variance of the residual error, $\var(\mathbf{J^F_{ss}} - \mathbf{J^C})$, matches well
the variance of the non-systematic error term, $\boldsymbol{\Gamma}$ in (\ref{eq:DefGamma}) even for stronger values of $J_0$
and only diverging for the systems of both strong couplings ($J_0=1$) and large hidden populations
(fraction visible $<0.5$).
This trend can be seen also in Figure \ref{fig:phasedaigrams} where the relative errors have been plotted, normalized by
the variance of the couplings for systems of varying coupling strength and fraction visible. Here the regime where the 
correction can no longer improve the inference (top left corner) appears as a phase transition in the space of coupling strength and 
fraction observed. 
Importantly, in the regime where the correction can be useful, the error due to subsampling
quickly becomes significant, as illustrated in Figure \ref{fig:undersampling}, surpassing the expected error 
due to data length limitations at reasonably small data lengths.

\section{Discussion}
\label{sec:Results}

In this paper we deal with the
effects of couplings and states of the hidden variables in a mean-field type approximation on the inference of the couplings 
between the observed nodes. We showed that the hidden population provides two sources of errors. 
The first term can be largely attributed to the underestimated size of the system, 
while the second term takes into account the hidden to observed correlations and couplings. 
Neither of the two contributions to the error (\ref{eq:approximationerror}) depends explicitly on interactions between hidden units that 
introduce higher order corrections to the learning. 
The reconstruction of the visible part of the network is indeed basically unaffected when removing the hidden to hidden 
connections \cite{dunn2013learning}, motivating the development of algorithms that neglect them \cite{battistin2015belief}\cite{tyrcha2014network}.  

We showed that in case of weakly interacting units, the non-systematic error from ignoring the hidden
population is small with respect to the strength of the couplings. 
In the weak couplings regime the non-systematic term can then be safely neglected and the systematic one treated using a mean field approximation. 
This allowed us to integrate the latter in the inference (equation (\ref{eq:rotationequation})),
resulting in a significantly better estimate of the true couplings, even for more strongly coupled systems.
For networks with strong couplings, however, and, in particular,
those with comparatively large hidden populations, the non-systematic error component dominates and could not be well addressed following this approach.

As mentioned earlier, in many data sets we are forced to ignore the large majority of the
relevant network. Attempting to make conclusions from parameters inferred from such a tiny fraction seems ridiculous
but the results here would suggest that, at least for weakly coupled systems, there might be hope.
The question then begs itself, \emph{how strongly connected are the systems we care about?}
If strongly connected, as very well could be the case \cite{schneidman2006weak}, then we may have reason to be concerned. 

An alternative strategy to the reconstruction of partially observed networks, that also avoids addressing the dark side directly, 
would be to learn an external field in addition to the connectivity \cite{borysov2015us,bury2013statistical,zeng2014financial}. 
The external field would then \textit{explain away} correlations between observed nodes driven by hidden nodes, preventing the 
observed to observed couplings to adjust for those. 
In this paper we intentionally do not include the external field in the model (\ref{eq:transitionIsing}), 
even though the dynamical model allows for it. 
We indeed expect that, to some extent, the inferred external field will take care of the input from hidden to observed units, 
reducing the error in the reconstructed connectivity \cite{kulkarni2007common}.  
As an example, by learning time-varying fields that explain away correlations in the data due to aspects of the experiment, 
the authors in \cite{dunn2015correlations} were able to infer seemingly significant functional connectivity from a very small subpopulation of neurons.
Further work should shed light on the extent to which the effects of the hidden population can be accounted for by including 
different non-stationary external fields.

\section*{Acknowledgments}
We thank Yasser Roudi for his substantial contribution to the original ideas behind this manuscript and useful discussions during the preparation. We are grateful to the Referees, for carefully reading the manuscript and constructively sharing their comments on the topic with us.  
This work has been funded by the Kavli Foundation and the Norwegian Research Council Centre of Excellence scheme.


\bibliographystyle{abbrv}
\bibliography{allreferences}

\end{document}